\documentclass[12pt]{iopart}

\usepackage{graphics}
\usepackage{graphicx}
\usepackage{epsfig}
\begin{document}

\title[Detection Strategies for Extreme Mass Ratio Inspirals]{Detection Strategies for
Extreme Mass Ratio Inspirals}

\author{Neil J. Cornish}

\address{Department of Physics, Montana State University, Bozeman,
MT 59717, USA}

\begin{abstract}
The capture of compact stellar remnants by galactic black holes provides a unique laboratory for
exploring the near horizon geometry of the Kerr spacetime, or possible departures from general
relativity if the central cores prove not to be black holes.
The gravitational radiation produced by these Extreme Mass Ratio Inspirals (EMRIs) encodes a detailed
map of the black hole geometry, and the detection and characterization of these signals
is a major scientific goal for the LISA mission. The waveforms produced
are very complex, and the signals need to be coherently tracked for hundreds to thousands of cycles
to produce a detection, making EMRI signals one of the most challenging data analysis problems
in all of gravitational wave astronomy. Estimates for the number of templates required to perform
an exhaustive grid-based matched-filter search for these signals are astronomically large, and
far out of reach of current computational resources. Here I describe an alternative approach
that employs a hybrid between Genetic Algorithms and Markov Chain Monte Carlo techniques, along
with several time saving techniques for computing the likelihood function. This approach has
proven effective at the blind extraction of relatively weak EMRI signals from simulated LISA
data sets.
\end{abstract}


\submitto{\CQG}

\maketitle

\section{Introduction}

The capture of stellar remnants - white dwarfs, neutron stars and black holes - by the massive
black holes that are thought to reside at the centers of most galaxies provide an excellent
laboratory for performing precision tests of general relativity. The large discrepancy in
the masses allows the smaller body to be treated as a perturbation to the spacetime
of the galactic black hole, and the evolution of the system can be treated analytically.
The gravitational wave signals from these systems encode effects such as frame dragging,
periastron advance, and spin-orbit coupling in highly modulated waveforms. The detection and
characterization of these Extreme Mass Ratio Inspirals (EMRIs) is a key
science goal of future space based gravitational wave detectors such as LISA~\cite{LISA}. Finding EMRI
signals in the output of a noisy detector presents a challenging data analysis problem as the signals
have to be followed for tens of thousands of cycles in order to accumulate sufficient signal-to-noise
ratio (SNR) for detection. It has been estimated~\cite{gair} that it would take of order $10^{40}$
templates to perform an exhaustive matched-filter search for these signals. We either have to hope
for a major advance in computing, or look for sub-optimal techniques to the EMRI detection problem.

It is natural to consider hierarchical strategies that either work with some subset of the
data (for example, smaller time segments), or coarser parameter search grids that are then
refined in the regions surrounding candidate detections. In Ref.~\cite{gair} a stack-slide~\cite{stack}
search algorithm was put forward that combines both of these strategies, and it was estimated that
with year 2013 computing resources it would be possible to detect systems with SNRs greater than 30,
which is a factor of two or three worse than could be done with a fully coherent search. Implementing
the stack-slide algorithm is a non-trivial task, but it would be interesting to see how it performs
on the simulated EMRI data sets that have been produced for the Mock LISA Data Challenges
(MLDCs)~\cite{Arnaud:2006gm}. Another approach to the EMRI detection problem is to use time-frequency
techniques~\cite{tf} to search for tracks in spectrograms of the data. This approach has the
advantage of being computationally cheap, and it has been applied with some success to the
MLDC data sets~\cite{tfmldc}, but there are concerns about how it will perform when applied to
more realistic data containing the signals from millions of galactic binaries, multiple massive
black hole binaries and hundreds of EMRIs.

Here I describe a collection of techniques, that when combined, have proven effective at detecting
blind EMRI signal injections in simulated LISA data. The techniques fall into two categories:
the first are computational techniques that either ``soften up'' the likelihood function or
speed up its calculation; the second are stochastic search techniques that facilitate the
efficient exploration of high dimensional spaces. The former are equally applicable to traditional
grid based searches, while the latter provide a more efficient search strategy.

The search strategy uses a hybrid of techniques from Markov Chain Monte Carlo and Genetic Algorithms.
Key elements of the algorithm are the use of multiple chains running at different ``temperatures'',
with communication between chains via Metropolis-Hastings exchange and genetic cross-over. The
individual chains are updated by a mixture of techniques, including jumps along Eigendirections of
the Fisher Information Matrix, Differential Evolution, mode jumping proposals between 
harmonics of the waveform, and hill climbing moves (using conjugate gradient or Nelder-Mead).
While this may sound like a ``Kitchen Sink Algorithm'', the various elements are not thrown
together haphazardly: there is a method to the madness.  The individual techniques were developed
by researchers in many fields of science (only the implementation of mode jumping described here is new),
and almost all of the techniques have been used at one time or another in LISA data analysis
studies (MCMC, Tempering, and Fisher Matrix based jumps were first introduced in Ref.~\cite{Cornish:2005qw};
Genetic Algorithms and Nelder-Mead in Ref.~\cite{Crowder:2006wh}; mode jumping in
Refs.~\cite{Crowder:2006eu,Cornish:2006ms}; and multiple tempered chains in
Ref.~\cite{Littenberg:2009bm}). Many of these techniques have also been studied in the context
of LIGO-Virgo data analysis (see {\it e.g.} Refs.~\cite{Christensen:1998gf,
Rover:2006bb,Lightman:2006rp,vanderSluys:2008qx}). More important than the specific techniques
being used are the basic principles on which the algorithm is based. The modular structure
of the algorithm makes it easy to incorporate new techniques as they become available.
A closely related approach to EMRI detection that shares a common heritage to the approach
outlined here is described in Ref.~\cite{Babak:2009ua}.

I begin in Section~\ref{sec:wavelike} with a brief description of the EMRI waveforms
used in this study and the techniques used to soften the likelihood surface and speed
up the calculation. Then in Section~\ref{sec:search} I describe the design of the search
algorithm and the new approach to mode jumping.

\section{Waveforms \& Likelihood}\label{sec:wavelike}

Waveform models describing the inspiral of a small compact object into a much more massive
Kerr black hole have come along way in the past decade, but we are still lacking a complete
description. For the purposes of data analysis development, the MLDC taskforce adopted the
philosophy of Barack \& Cutler~\cite{bc}, who argued that most of the features
of the full analysis could be captured by considering ``kludge waveforms'' that are parameterized
by the same 14 parameters as the exact model, and share the same qualitative features such as
eccentricity, periastron precession, spin-orbit induced precession of the orbital plane and
the loss of energy and angular momentum due to gravitational wave emission. On the other hand,
the Barack-Cutler kludge waveforms fail to capture relativistic effects such as ``zoom-whirl''
behavior~\cite{Glampedakis:2002ya} and non-adiabatic resonances~\cite{Flanagan:2010cd}, so developing
a data analysis algorithm that is capable of detecting these simpler signals is only a first step
towards tackling the EMRI detection problem.

A full description of the Barack-Cutler waveform model used in the MLDC can be found in Ref.~\cite{mldc3}.
The waveforms are described at any instant by a collection of harmonics $f_{nkm}$ of the azimuthal orbital
frequency $f_\Phi=\nu$, the periapse precession frequency $f_{\tilde \gamma}$ and the orbital plane precession
frequency $f_\alpha$:
\begin{equation}
f_{nkm} =  n f_\Phi + k f_{\tilde \gamma} + m f_\alpha \, ,
\end{equation}
where $n,k,m$ are integers. Unlike the full EMRI waveforms, the kludge waveforms are limited to
harmonics with $k=2$ and $m \in [-2,2]$ (though this is from obvious in the precessing frame
of Ref.~\cite{bc}), and the relatively low eccentricity of systems considered
in the MLDC data sets allows us to ignore harmonics with $n > 6$. Since $f_\alpha$ is roughly an
order of magnitude smaller than $f_\Phi$ or $f_{\tilde \gamma}$, the harmonics come in bands of five labeled by
$n$. The frequency and amplitude
of each harmonic evolve slowly over time as the system loses energy and angular momentum. The LISA
response function imparts additional modulations on the waveform due to the motion of the detector
about the Sun. The codes used to generate the EMRI waveforms for the MLDC data sets do not exploit
the separation of timescales available in the problem, and as a result each waveform takes seconds
to minutes to generate on the processors available today. 

When a waveform model is available, gravitational wave data analysis boils down to comparing a model
waveform $h(t,\vec{\lambda})$, described by parameters $\vec{\lambda}$, to the observed time series
$s(t)$. The goodness of fit is measured by the chi-squared $\chi^2 = (s-h \vert s-h)$, where $(\cdot \vert \cdot)$
denotes the noise weighted inner product. For Gaussian noise the likelihood $p(s\vert \vec{\lambda})$
is proportional to $\exp(-\chi^2/2)$. Calculating the waveforms and noise weighted products directly
using the tools provided by the MLDC taskforce can cost minutes per point in parameter space, and
with a complicated 14 dimensional parameter space to explore, the run time of a search can be
prohibitive. Thus, the first priority is to find ways to speed up the calculation of the likelihood,
and if possible, reduce the dimension of the search.


It is possible to dramatically speed up the likelihood calculation using variants of the techniques described
in Refs.~\cite{Cornish:2010kf, Cornish:2007if}. The first step is to heterodyne the data using a collection
of trial frequency evolution functions, $f_n(t)$, for each set of harmonics. The heterodyned data is then
low pass filtered,
which effectively throws away all the data that is not close to the $f_n(t)$ evolution track (as viewed in a
time-frequency representation). For the MLDC data sets most of the power is concentrated in the $n=2,3$ bands,
and these are usually sufficient for detection purposes. To achieve an effective
compression of the data we need $f_n(t)$ to approximately follow the evolution of the $f_{n20}(t)$ harmonic,
which depends on parameters such as the mass and spin of the central black hole, and the initial orbital
frequency. To account for this, a coarse grid of heterodyne functions are used to generate multiple
reduced data sets that are stored for later use. The template parameters $\vec{\lambda}$ are used
to select the appropriate reduced data set for the likelihood calculation. Since the heterodyning
procedure takes out most of the frequency evolution, the EMRI signal now looks like a collection of
galactic binaries (albeit ones that have a slowly changing amplitude and polarization), and it is possible to
generate the individual harmonics directly in the Fourier domain using a slightly modified version of
the fast galactic binary generator described in Ref.~\cite{Cornish:2007if}. The net result is a
likelihood calculator that runs thousands of times faster than the direct approach.

\begin{figure}
\begin{center}
\includegraphics[width=4.5in]{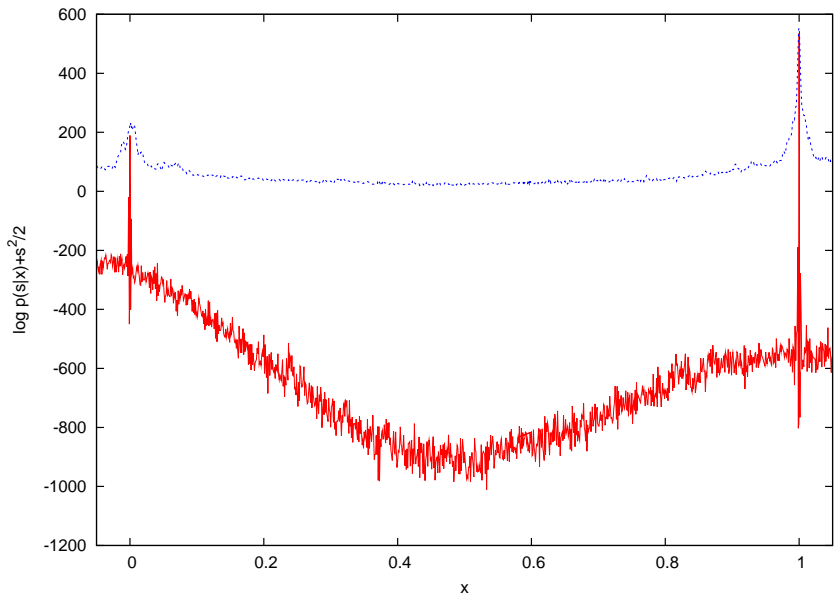}
\end{center}
\caption{Cross sections of the likelihood function along a line in parameter space connecting
a secondary mode to the primary mode for EMRI \# 3.3.2 from Round 3 of the MLDCs (The secondary mode
corresponds to a harmonic mis-match) The solid (red) line shows the standard log likelihood, while
the dotted (blue) line shows the analytically maximized log likelihood.}
\label{like}
\end{figure}

In addition to speeding up the likelihood evaluations it is also possible to reduce the effective
dimension of search space by analytically maximizing the likelihood with respect to some of the
waveform parameters. Working in the Fourier domain it is possible to maximize the likelihood by
analytically rotating the phase of the waveform and sliding it in time. The corrected distance
$D'$ to the EMRI system can be found via the relation $D' = D \, (h\vert h)^{1/2}/(s\vert h)$.
The time-slide maximization is equivalent to solving for the initial orbital frequency
$\nu_0 = f_\Phi(0)$, while the phase maximization of three or more harmonics fixes the
initial phases $\Phi_0$, $\tilde{\gamma}_0$ and $\alpha_0$. These standard analytic maximization
techniques take care of 5 of the 14 EMRI parameters. Not only does the maximization shrink the
search space, it also ``softens up'' the likelihood surface: suppose that we have a fairly good
match with some set of parameters
$\vec{x}$, and the algorithm proposes a move to a new set of parameters $\vec{y}$ with slightly different
mass and spin parameters, and with the phase parameters held fixed. In general the likelihood at
$\vec{y}$ will be lower than the likelihood at $\vec{x}$ since the frequency and phase of the waveform
shifts as the parameters are changed. But if we maximize over the initial frequency and phase these
shifts can be compensated for, and there is a good chance that the modified set of parameters $\vec{y}^{\, '}$
will return a likelihood comparable to that at $\vec{x}$. Figure~\ref{like} shows cross sections of the
regular and maximized log likelihood along a line in parameter space connecting a secondary mode
to the primary mode. The valley between the peaks is far shallower, and the peaks are far wider for
the maximized likelihood, which makes it easier for a search algorithm to finds its way from one peak
to the next.

The softening of the likelihood can be taken a step further if we allow for additional flexibility in
the maximization. For example, while the physical harmonics have relative phases that are set by
$\Phi_0$, $\tilde{\gamma}_0$ and $\alpha_0$, there is nothing stopping us from phase maximizing each
of the harmonics separately during the search. This additional flexibility can help compensate for
incorrect values of the other signal parameters. Similarly, while there can only be one value for
$\nu_0$, it is possible to introduce some additional flexibility into the time-shift maximization step.
Separately maximizing each harmonic will not work since all of the harmonics will slide over to match
the brightest harmonic of the signal, but it is possible to perform a simultaneous maximization that
maintains a frequency separation between the $f_\alpha$ harmonics. This allows us to solve for
$f_\alpha(0)$ in addition to $\nu_0$, the combination of which provide an improved estimate for
the spin of the larger black hole. Additional softening can be achieved by dividing the signal up
into shorter time segments and phase maximizing each harmonic in each segment individually.
The price that has to paid for softening the likelihood is that the templates can more easily latch
onto features in the instrument noise. Ultimately this is not a serious problem since the maximization
is turned off as the algorithm switches from the initial search phase to a MCMC exploration of the
posterior distribution function, and matches that only exist because of the maximization are lost.

\section{Search Algorithm}\label{sec:search}

Our goal is to detect and characterize EMRI signals. The detection phase is an optimization problem -
we seek to find the maximum of the posterior distribution function $p(\vec{\lambda} \vert s) =
p(\vec{\lambda}) p(s \vert \vec{\lambda})/p(s)$. Here $p(\vec{\lambda})$ is the prior distribution,
$p(s \vert \vec{\lambda})$ the likelihood, and $p(s)$ is the marginal
likelihood or evidence. The characterization phase is where we map out the peaks of the posterior
distribution and compute confidence intervals for the recovered parameters.

The characterization stage is now well understood within the gravitational wave community, and
techniques such as Markov Chain Monte Carlo and Nested Sampling have been shown to do a very
good job of mapping the peaks of the posterior distribution, including those for EMRI signals.
Finding the modes, or peaks, is another problem entirely, and while there is a vast literature
devoted to similar optimization problems spanning many fields of study, there is no
universal optimization algorithm that outperforms all others in every situation.

The MCMC algorithms used to map the posterior can perform quiet well as search algorithms. Indeed,
there are theorems~\cite{tierney} which prove that samples from a Markov Chain will always converge
to give the posterior distribution (which in practice means that the chains will end up spending most
of their time exploring the peaks of $p(\vec{\lambda} \vert s)$). However, the rate of convergence to
the target distribution (``burn-in time''), and the number of samples needed to accurately reconstruct
the posterior (``mixing time'') depend on the particular implementation of the Metropolis Hastings algorithm
being used, and on the nature of the target distribution. For a wide class of algorithms it is possible to
prove that the Markov chains produced by the Metropolis-Hastings algorithm are
{\em geometrically ergodic}~\cite{meyn}, and for such chains there are theorems that provide bounds on the
burn-in and mixing times - see Ref.~\cite{jones} for an accessible review. Unfortunately these bounds are
very weak, and the number of iterations required for the burn-in phase can be impractically large.

From an optimization standpoint, MCMC algorithms are not sufficiently ``greedy''. To be Markovian the
chains must respect detailed balance, that is, steps in either direction along the chain must be
equally probable. But that is the last thing you want in a search, where the goal is to go uphill.
On the other hand, simple hill climbing schemes such as conjugate gradient or Nelder-Mead methods
have a tendency to climb up the first bump they find and get stuck. There is a clear
advantage in combining the tendency to go up hill ({\it e.g.} Nelder-Mead) with the ability to go
down hill (MCMC). Wider exploration of the parameter space can be promoted by tempering the
likelihood surface, which is done by modifying the likelihood function: $p(s \vert \vec{\lambda}) \rightarrow
p(s \vert \vec{\lambda})^{1/T}$, where $T \geq 1$ plays the role of temperature. In the limit
$T \rightarrow \infty$ the likelihood function is flat, and the Markov Chain recovers the prior
distribution. Simulated annealing schemes can be successful, whereby the temperature starts at a
high value and is steadily reduced to $T=1$ via some cooling schedule, but the performance can be quiet
sensitive to the choice of cooling scheme. A more robust alternative is Parallel Tempering~\cite{ptmcmc},
whereby multiple chains are run in parallel, each with a fixed temperature $T_i$ on a temperature ladder
$T_1=1 < T_2 < T_3 \dots$. The parameters of adjacent chains on the temperature ladder can be
exchanged through a Metropolis-Hastings move, which results in solutions with higher likelihood sinking
down to the lowest rung on the temperature ladder. The cold chains help the search to ``remember'' good
solutions, while the hot chains are free to explore the full parameter space and find other modes
of the posterior. The effectiveness of Parallel Tempering is affected by the spacing of the chains
on the temperature ladder and the number of chains used, but these parameters are easy to tune.

A Parallel Tempered MCMC (PTMCMC) algorithm shares many attributes with Evolutionary or Genetic Algorithms
(for descriptions of Genetic Algorithms in the context of gravitational wave data analysis see
Refs.\cite{Crowder:2006wh,Lightman:2006rp,Gair:2009cx,Petiteau:2010zu}).
In PTMCMC there are a population of ``organisms'' (the individual chains), each with its own ``DNA'' (the
parameter values). Mutation of the DNA occurs by Metropolis-Hastings updates of each chain, and the temperature
ladder acts like a selection mechanism with the ``fitter'' organisms sinking to the bottom,
where their DNA is more likely to be preserved. The key ingredient that is missing is reproduction or
crossover. In PTMCMC the chains can swap parameters, but their is no mechanism for mixing or combing the
parameters from different chains. This deficiency can be corrected by allowing genetic crossover between
the chains, leading to an algorithm called Evolutionary Monte Carlo (EMC)~\cite{EMC1,EMC2}. The
EMC algorithm preserves detailed balance in the genetic crossover operation, so the resulting samples are
those of a Markov chain, but this severely limits the types of reproduction that can be used.
During the search phase we are not concerned about reversibility, and it is more effective to use
crossover operations that violate detailed balance, such as intermediate and line recombination~\cite{reco}.
Another way to view the EMC algorithm is as a Genetic Algorithm the uses Metropolis-Hastings updates
for the mutation operation, which has proven to be far more effective than the random binary coded mutation schemes
used in Refs.~\cite{Crowder:2006wh,Gair:2009cx,Petiteau:2010zu}.

In any MCMC scheme the choice of proposal distribution, from which new candidate points are
drawn, is crucial to the success of the algorithm. Past experience has shown that it helps to have
a mixture of proposal distributions, and the current version of our EMC algorithm uses a combination of
Fisher Matrix jumps along eigendirections of the Fisher Information
Matrix $\Gamma_{ij} = -\langle \partial_i \partial_j \log p(s \vert \vec{\lambda})\rangle =
( h_{,i}\vert h_{,j})$, Conjugate Gradient jumps along the gradient of the likelihood,
$\nabla \log p(s \vert \vec{\lambda})$~\cite{CGMC}, draws from the prior $p(\vec{\lambda})$, small
Gaussian jumps in each parameter, Differential Evolution (DE)~\cite{DEMC}, and mode hoping jumps based
on the EMRI harmonics. The DE moves are a new addition to the mix, and they are
extremely effective during the characterization phase. The DE procedure is to propose a jump
from the current location to a new point along a vector connection two points drawn from the past
history of the chain. This technique is very well suited to exploring posterior distributions with
highly correlated parameters, and is asymptotically Markovian in the limit where the past history
of samples is large. During the search phase it is better to use local DE, where the jump
direction is drawn from the recent history of the chain.

\begin{figure}
\begin{center}
\begin{tabular}{cc}
\includegraphics[width=2.5in]{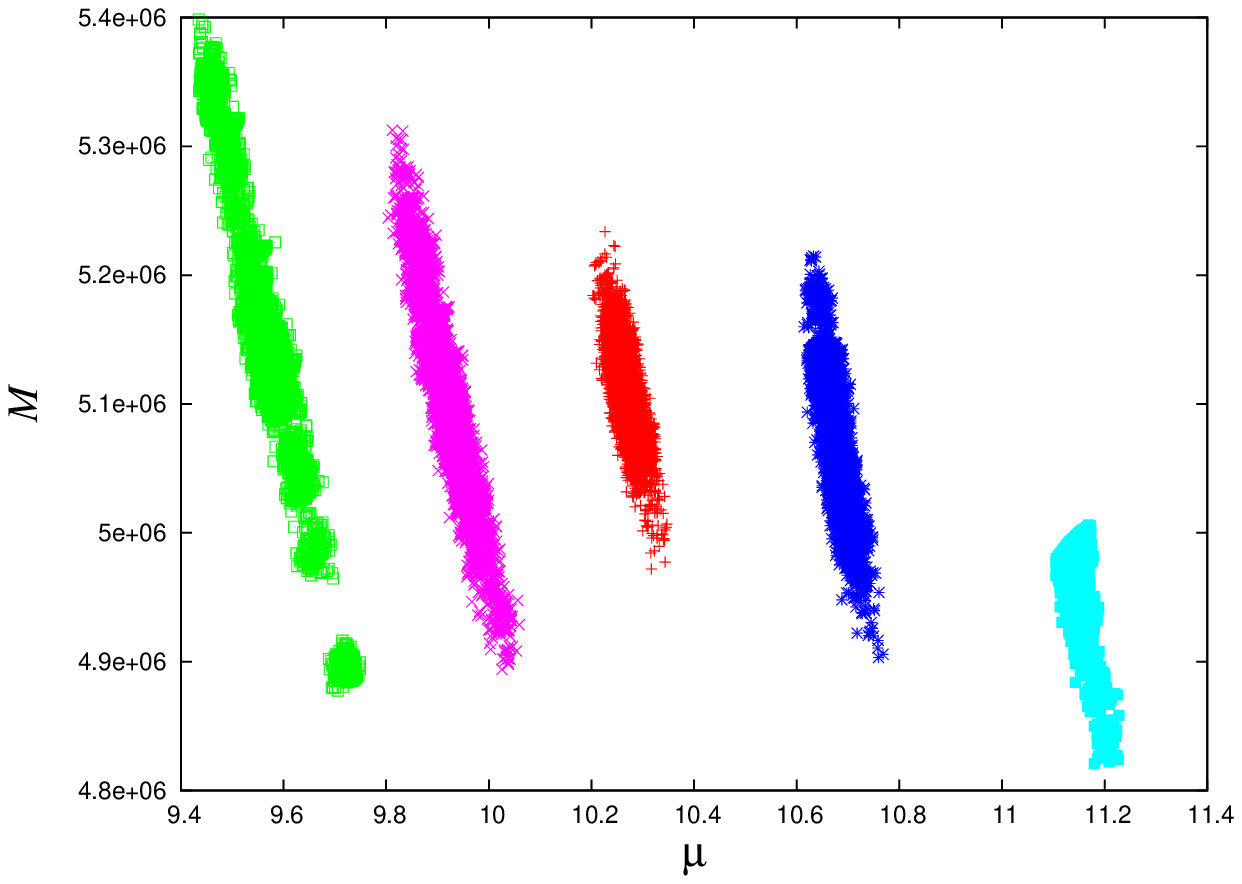} & 
\includegraphics[width=2.5in]{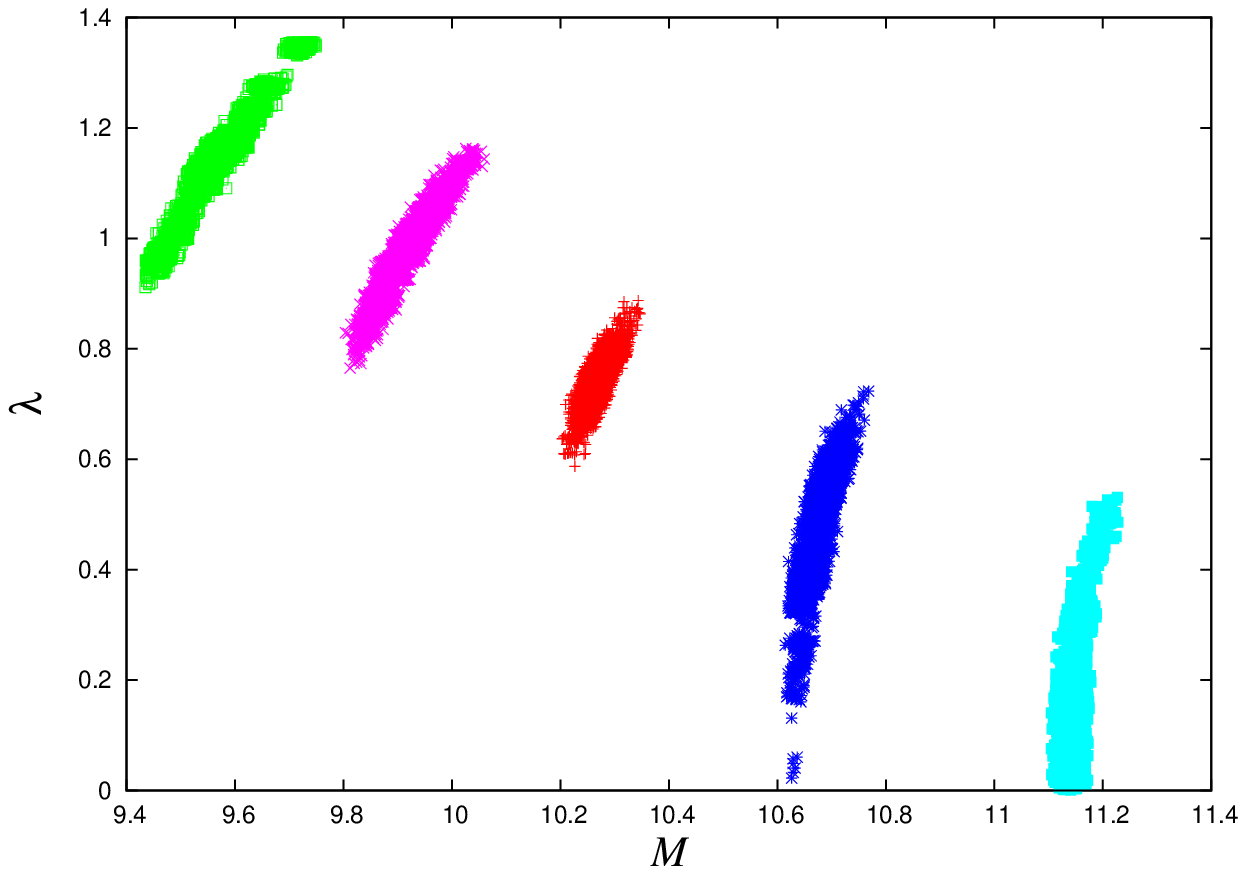} \\
\includegraphics[width=2.5in]{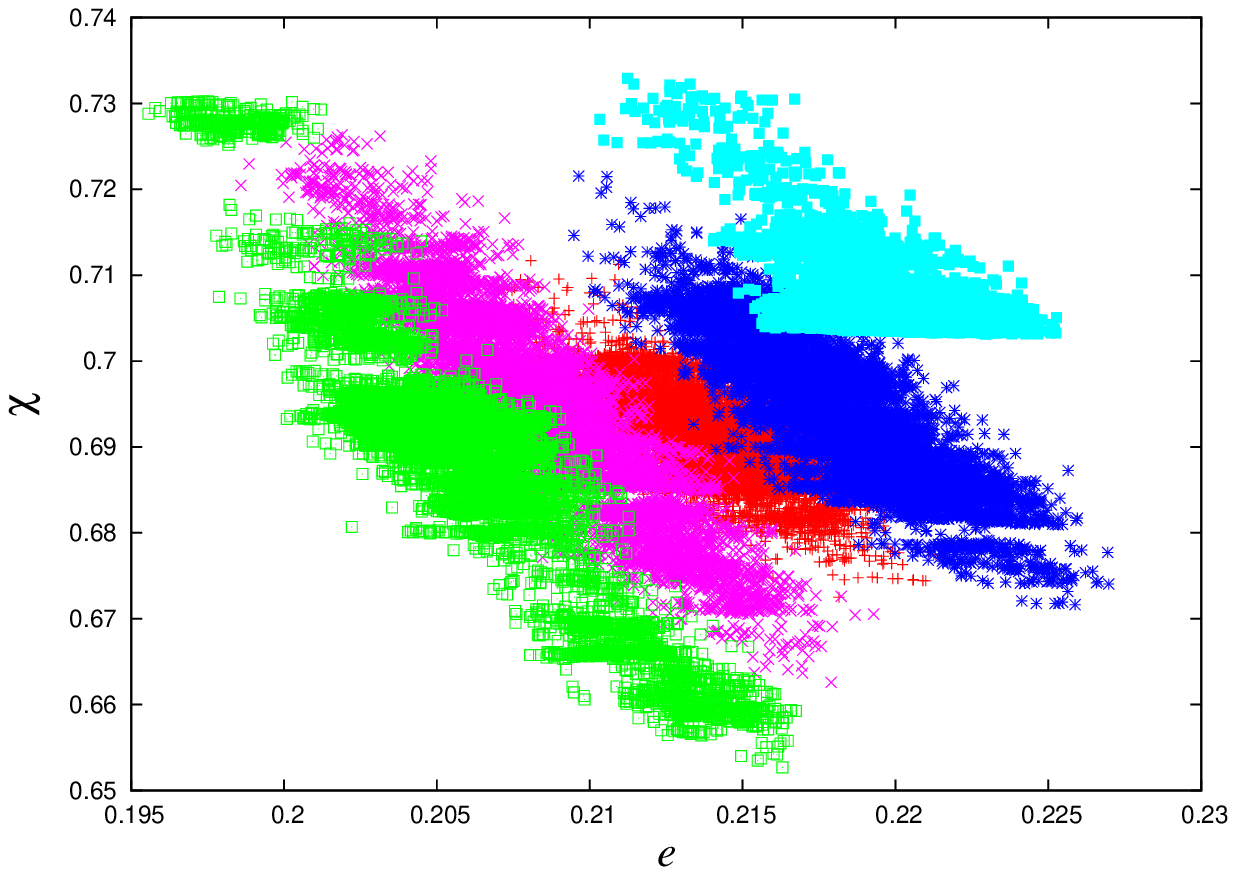} &
\includegraphics[width=2.5in]{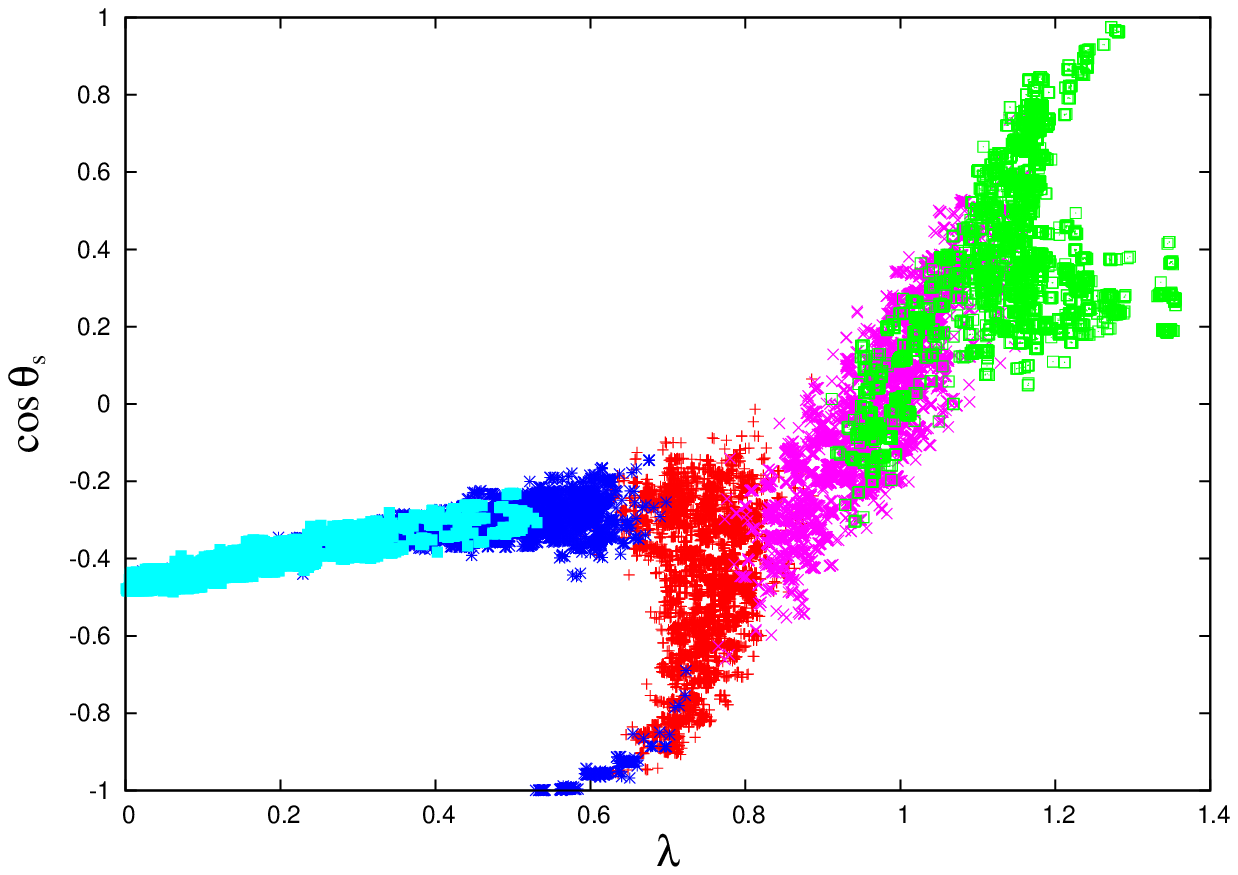}
\end{tabular}
\end{center}
\caption{Scatter plots from the EMC chains used to find the parameter values that yield harmonic
jumps. The color coding is green of $j=2$, magenta for $j=1$, red for $j=0$, blue for $j=-1$ and
cyan for $j=-2$.}
\label{hjumps}
\end{figure}

The mode hoping jumps address one of the biggest problems encountered by earlier versions of our
EMRI search algorithm. In Round 1B of the MLDC~\cite{:2008sn}, the Montana group's entries for
data sets 1B.3.2 and 1B.3.3 corresponded to secondary modes of the posterior, with $f_{\tilde \gamma}$
offset from the correct value by $\pm f_\alpha/2$. This offset seriously biased the recovered mass and
spin parameters. The latest version of the algorithm turns the harmonic mismatch problem from
a detriment to a benefit, as now whenever the search locks onto a secondary mode of the posterior
corresponding to a harmonic mismatch, it quickly transitions to the primary mode by accepting a mode
hoping proposal. The mode hops work as follows. Suppose the chain is at $\vec{x}$ with harmonics
$f_{n2m}(\vec{x},t)$. We now propose a jump to a new location $\vec{y}$ with harmonics
$f_{n2(m+j)}(\vec{y},t) \approx f_{n2m}(\vec{x},t)$, where $j$ can equal $-2,-1,1,2$. The trick is in finding the
parameters $\vec{y}$ that yield the desired harmonic offset. This is done using an EMC search
of a likelihood defined by the chi-squared
\begin{equation}
\chi^2 = \alpha  \sum_{n,m} \sum_{k=0..2} \left(\frac{d}{dt}\right)^k\left(f_{n2m}(\vec{x},t)-f_{n2(m+j)}(\vec{y},t)\right)\Big|_{t=0} \frac{T^{k-1}}{k !} \, .
\end{equation}
Here $T$ is the time of observation and $\alpha$ is a scaling parameter that sets the tolerance for the search.
Physically, this chi-squared measures the frequency offset of the harmonic tracks using a Taylor series
approximation. Because this chi-squared is very cheap to compute, it only takes a fraction of a second to
find a good solution for $\vec{y}$. With the harmonic jumps in the mix, the search no longer gets stuck
at secondary maxima caused by harmonic mis-match. Figure~\ref{hjumps} shows 2-dimensional scatter plots
of the parameters $\vec{y}$ found by the harmonic jump EMC chains.

In Round 3 of the MLDC, an earlier version of the EMC algorithm successfully detected 3 of the 5
moderate SNR, overlapping EMRI signals that had been injected into a single data set~\cite{Babak:2009cj}.
The two that were missed had lower mass central black holes, and consequently higher frequencies.
Since the earlier version of the search code did not heterodyne the data, the computational cost
of searching for these low mass systems was higher, and the search had not converged to the
correct solution by the challenge deadline. The new and improved version of the code now finds
all the overlapping EMRI signals with no difficulty, as will be demonstrated in the next
round of blind challenges.

\section*{Acknowledgments}
It has been my pleasure to collaborate with Edward Porter and Jonathan Gair on the development
of several of these EMRI detection techniques. I have also benefited from discussion with
Curt Cutler, Stas Babak and Michele Vallisneri.

\end{document}